\documentclass[journal=jpccck,manuscript=article,layout=twocolumn]{achemso}

\usepackage{chemformula} 
\usepackage[T1]{fontenc} 
\SectionNumbersOn
\usepackage{siunitx}
\usepackage{bm}%
\usepackage[colorlinks=true,linkcolor=blue]{hyperref}%
\usepackage{graphicx}
\usepackage[utf8]{inputenc}
\usepackage{epstopdf}
\usepackage{booktabs}
\usepackage{abstract}
\usepackage{lineno}

\expandafter\ifx\csname package@font\endcsname\relax\else
 \expandafter\expandafter
 \expandafter\usepackage
 \expandafter\expandafter
 \expandafter{\csname package@font\endcsname}%
\fi
\author{Jan Weinreich}
\affiliation{Universit\"{a}t G\"{o}ttingen, Institut f\"{u}r Physikalische Chemie, Theoretische Chemie, Tammannstra\ss{}e 6, 37077 G\"{o}ttingen, Germany}
\author{Mart\'{i}n Leandro Paleico}
\affiliation{Universit\"{a}t G\"{o}ttingen, Institut f\"{u}r Physikalische Chemie, Theoretische Chemie, Tammannstra\ss{}e 6, 37077 G\"{o}ttingen, Germany}
\author{J\"{o}rg Behler}
\email{joerg.behler@uni-goettingen.de}
\affiliation{Universit\"{a}t G\"{o}ttingen, Institut f\"{u}r Physikalische Chemie, Theoretische Chemie, Tammannstra\ss{}e 6, 37077 G\"{o}ttingen, Germany}
\altaffiliation{International Center for Advanced Studies of Energy Conversion (ICASEC), Universit\"at G\"ottingen, Tammannstra\ss{}e 6, 37077 G\"ottingen, Germany}
\title{Properties of $\alpha$-Brass Nanoparticles II: Structure and Composition} 

\let\oldmaketitle\maketitle
\let\maketitle\relax
\begin{document}

\twocolumn[
\begin{@twocolumnfalse}
\oldmaketitle
\begin{abstract}
Nanoparticles have become increasingly interesting for a wide range of applications, because in principle it is possible to tailor their properties by controlling size, shape and composition. One of these applications is heterogeneous catalysis, and a fundamental understanding of the structural details of the nanoparticles is essential for any knowledge-based improvement of reactivity and selectivity. 
In this work we investigate the atomic structure of brass nanoparticles containing up to 5000 atoms as a typical example for a binary alloy consisting of Cu and Zn.
As systems of this size are too large for electronic structure calculations, in our simulations we use a recently parametrized machine learning potential providing close to density functional theory accuracy. This potential is employed for a structural characterization as a function of chemical composition by 
various types of simulations like Monte Carlo in the Semi-Grand Canonical Ensemble and simulated annealing molecular dynamics.
Our analysis reveals that the distribution of both elements in the nanoparticles is inhomogeneous, and zinc accumulates in the outermost layer, while the first subsurface layer shows an enrichment of copper. Only for high zinc concentrations alloying can be found in the interior of the nanoparticles, and regular patterns corresponding to crystalline bulk phases of $\alpha$-brass can then be observed. The surfaces of the investigated clusters exhibit well-ordered single-crystal facets, which can give rise to grain boundaries inside the clusters. The melting temperature of the nanoparticles is found to decrease with increasing zinc-atom fraction, a trend which is well-known also for the bulk phase diagram of brass.
\end{abstract}
\end{@twocolumnfalse}
]

\section{Introduction} \label{sec:intro}

Metal and alloy nanoparticles (NPs) have found applications in numerous fields\cite{P3458}, one of the most prominent being heterogeneous catalysis\cite{P5338,P5542,P5543}. Brass, a commercially important alloy of Cu and Zn, is a textbook example for binary alloys, and its bulk properties have been thoroughly studied for decades~\cite{P2119,P2156,P4405,p5553,P5704,brass,P5705,P5706}. For low Zn contents ($< 40~\%$) brass adopts the face-centered-cubic (FCC) crystal structure with varying occupations of the lattice sites by Cu and Zn\cite{brass} called the $\alpha$-phase, which we will study in this work. In spite of detailed knowledge about the bulk phases of brass, little is known about brass NPs, and even though a lot of progress has been made in recent years, experimental synthesis and characterization of brass NPs remains a formidable challenge. Hence, to date only a few studies have addressed the synthesis of brass NPs~\cite{P5689,P5707,P5708,P5709,P5710}, and also theoretical studies are still rare~\cite{P5551,P5555,JanPaper1}.
\\
Although a variety of methods such as basin-hopping Metropolis Monte Carlo \cite{P5718,P5721,P5294} (BHMC), genetic algorithms\cite{P5722, P5713} and minima hopping\cite{P1065} are available to identify low-energy configurations of large metal NPs, the global optimization (GO) of these systems remains a challenging task \cite{P2789,P5632,doi:10.1002/qua.24462,PhysRevB.97.195424,doi:10.1063/1.3077300, C9CP06967D}, because alloy NPs have many degrees of freedom such as size, shape, composition and distribution of the elemental species\cite{P5699}. 
\\
As the system size increases, often many configurations with very similar energies are present, which is not only a challenge for the description of the potential energy surface (PES) but also creates a situation where the ensemble behavior is more relevant to the properties of the system than a single global minimum energy structure. At this point the problem generalizes to generating a wide range of relevant structures and analyzing average properties of the generated ensemble. Equally important is the role of non-zero temperature, which also results in the population of many different low-energy states of the system. 
\\
For this purpose, we study NPs under Semi-Grand Canonical Ensemble \cite{P5442,P5443,P5715} (SGCE) conditions where the number of atoms is constant but the copper-to-zinc ratio can change allowing to search for the lowest energy distribution of elements and the optimal composition as demonstrated in several previous applications \cite{Calvo2792,Atanasov2009, P5716, P5717}.
\\
Simulations of large NPs crucially depend on the fast evaluation of the PES for a large number of configurations. To address this issue many types of empirical potentials~\cite{EAM,EAM2,P0820,P0819,P0751} have been proposed and parameterized to density functional theory (DFT) data~\cite{P2786,P5726,P5723, P2785,P5724,P5725}.
In recent years in particular machine learning potentials have become a popular method to represent the PES with a high accuracy close to that of electronic structure methods \cite{P4885,P5673,P5793}, including high-dimensional neural network potentials (HDNNPs) \cite{P1174,P4444} as well as other neural network-based approaches~\cite{P5366,P5577,P5817,P4419}, Gaussian approximation potentials \cite{P2630}, moment tensor potentials \cite{P4862}, spectral neighbor analysis potentials \cite{P4644}, atomic cluster expansion
~\cite{P5794} and many others. All of these methods allow studying a large number of geometries containing thousands of atoms at computational costs comparable to empirical potentials. 
\\
In recent work it has been demonstrated that HDNNPs can be successfully applied to the GO and simulation of NPs \cite{P4470,C9CP00837C,P5899,P4028,P4474} beyond the system sizes and times scales which are accessible by DFT. Here, we employ a previously developed HDNNP for $\alpha$-brass \cite{JanPaper1} to carry out large-scale simulations and to characterize the structure and composition of large brass NPs containing up to about 5000 atoms. Our primary goal is to unravel the distribution of Cu and Zn at the surface as well as in the interior region of the NPs for different chemical potentials as a function of temperature.
\\
This article starts with a concise summary of the employed methods in Sec.~\ref{sec:methods} followed by a description of the details of the simulation protocols in Sec.~\ref{sec:comp_details}. The next section contains our results obtained in Monte Carlo simulations, employing fixed as well as relaxed atomic lattices (Sec.~\ref{sec:results_sgce}) to provide detailed insights into the low-energy configurations of brass NPs. Further, global structural changes, which are difficult to study by fixed-lattice methods, are investigated by performing simulated annealing (SA) simulations (Sec.~\ref{sec:sim_ann}). Our conclusions are drawn in Sec.~\ref{sec:conclusion}.

\section{Methods} \label{sec:methods}

\subsection{High-Dimensional Neural Network Potentials}

In the present work we use high-dimensional neural network potentials (HDNNPs) as introduced by Behler and Parrinello in 2007\cite{P1174} for the representation of the DFT potential energy surface (PES). It allows us to determine the energies and atomic forces of large brass nanoparticles as a function of the atomic positions with close to first-principles accuracy.
The method has been described in detail elsewhere,\cite{P4444,P4106,P5977} and here we give only a short summary. 
\\
In the HDNNP method, the total energy $E_{\rm tot}$ of a system containing $N$ atoms is constructed as a sum over individual atomic energy contributions $E_{i}$,
\begin{align}
   E_{\rm tot} = \sum_{i=1}^{N} E_{i}~. \label{eq:etot}
\end{align}
The atomic energy contributions depend on the respective local atomic environments that are defined by a cutoff radius $R_{\rm c}$, which is chosen to include a large number of neighboring atoms inside the resulting atomic cutoff spheres. 
The $E_i$ in Eq.~\ref{eq:etot} are obtained from individual atomic feed-forward neural networks. The inputs of these networks are vectors of many-body atom-centered symmetry functions\cite{doi:10.1063/1.3553717} (ACSFs), which are rotationally, translationally and permutationally invariant. The values of these ACSFs describe the positions of the neighboring atoms within the cutoff spheres and thus serve as structural fingerprints. For each element in the system, the atomic neural networks have fixed architectures, i.e. numbers of hidden layers and neurons per layer, and thus in the present case of brass there is one type of neural network yielding the energies of Cu atoms and a second one providing the energies of Zn atoms. Each of these networks is then evaluated as many times as atoms of the respective element are present in the structure of interest.
\\
The weight parameters of the atomic neural networks are determined iteratively to minimize the errors of the energies and forces in a reference set containing data from electronic structure calculations at the desired level of theory. This data set must cover the configuration space relevant for the intended simulations to ensure a reliable representation of the first-principles PES. 
Because the atomic energies and forces only depend on the local environments, a HDNNP can be trained using rather small systems~\cite{JanPaper1, doi:10.1021/acs.jctc.8b01288}, but then can be applied to much larger systems thus enabling large-scale simulations with the accuracy of the reference electronic structure method at a fraction of the computational costs.

\subsection{The Semi-Grand Canonical Ensemble \label{sec:theory_sgce}}

\begin{figure}
    \centering
    \includegraphics[width=\linewidth]{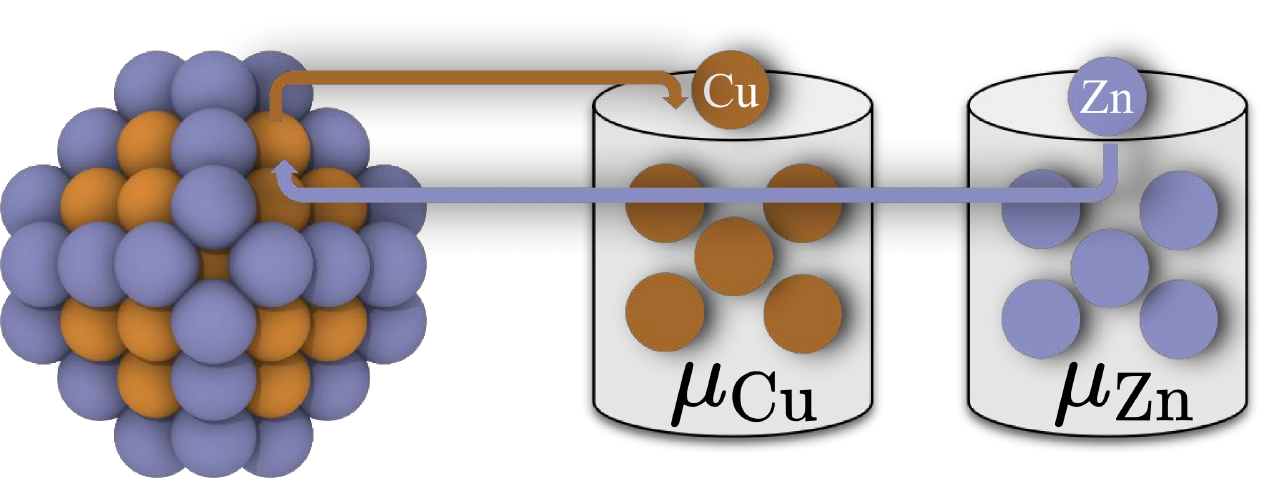}
        \caption{
        Simulations in the Semi-Grand Canonical Ensemble: Atoms in a brass cluster are exchanged with atoms from external particle reservoirs with chemical potentials $\mu_{\text{Zn}}$ and $\mu_{\text{Cu}}$, respectively. The chemical potential difference $\Delta \mu =\mu_{\text{Cu}}  - \mu_{\text{Zn}}$ controls the exchange of atoms (s. Eq.~\ref{eq:accprob}). Here a Cu atom is replaced by a Zn atom.}
    \label{fig:sgc}
\end{figure}

In simulations in the SGCE\cite{P5442,P5443,P5715} the stoichiometry of a system, i.e. the atom fractions of Zn and Cu, $x_{\text{Zn}}= \frac{N_{\text{Zn}}}{N_{\text{Zn}} + N_\text{Cu}}=\frac{N_{\text{Zn}}}{N}$ and $x_{\text{Cu}}= \frac{N_{\text{Cu}}}{N}$ with $x_{\text{Zn}}+x_{\text{Cu}}=1$, are allowed to change, while the number of atoms remains constant. The equilibrium atom fractions at a given temperature $T$ depend on the externally imposed chemical potentials $\mu_{i}$ for each element $i$. For a two element system like brass, the only relevant parameter is the difference between the assigned chemical potentials, $\Delta \mu =\mu_{\text{Cu}}  - \mu_{\text{Zn}}$. The system is connected to Cu and Zn particle reservoirs (s. Fig.~\ref{fig:sgc}), each at its given chemical potential, which can replace atoms already present in the system with atoms of a different element. This is in contrast with the Grand Canonical Ensemble, where the total number of atoms in the system can change.
\\
The effective energy function $U_{\text{SGCE}}$ in the SGCE is obtained as an extension of the structure-dependent potential energy $E_{\text{tot}}$ taking into account the external chemical potentials $\mu_{\text{Zn}}$ and $\mu_{\text{Cu}}$, weighted by the number of atoms of the respective element,

\begin{eqnarray}
U_{\text{SGCE}} &=& E_{\text{tot}} + N_{\text{Cu}} \mu_{\text{Cu}} +N_{\text{Zn}} \mu_{\text{Zn}} \nonumber \\
&=& E_{\text{tot}} + N x_{\text{Cu}} \Delta \mu + N \mu_{\text{Zn}}   ~.
\end{eqnarray}

A trial move in the SGCE changes the element of a randomly selected atom such that a Cu atom becomes a Zn atom or vice versa.
The Metropolis Monte Carlo (MMC)\cite{P0850} acceptance probability
\begin{eqnarray}
    P_\textrm{acc}  =& \textrm{min}& \left(1,e^{-\beta \Delta U_{\text{SGCE}}}  \right) \nonumber \\
    =& \textrm{min}& \left(1,e^{-\beta (\Delta E + \Delta \mu \cdot N \cdot \Delta x_{\textrm{Cu}})}  \right)~,
    \label{eq:accprob}
\end{eqnarray}
of the new structure depends on the difference $\Delta U_{\text{SGCE}}$ of the two structures such that the constant offset $N \mu_\text{Zn}$ cancels, with $\beta=1/k_{\textrm{B}}T$ and $k_{\textrm{B}}$ the Boltzmann constant.
Thus, not only the potential energy difference $\Delta E = E_{\rm tot,final} - E_{\rm tot,initial}$ after and before the trial move but also the term $\Delta \mu \cdot N \cdot   \Delta x_{\text{Cu}}$ influences the acceptance probability, such that the MMC algorithm samples configurations with a low value of $U_{\text{SGCE}}$. Effectively, the chemical potential term serves to bias the elemental composition, by compensating energy losses or gains when exchanging the element of an atom.

In summary, MMC simulations in the SGCE represent a useful tool to generate an ensemble of accessible configurations for given temperature and chemical potential difference of both elements. Note that we also include MMC swaps of atoms \textit{within} the system, without changing the ratio of Cu and Zn. In this case the exponent in Eq.~\ref{eq:accprob} simplifies to $-\beta \Delta E$, which is equivalent to a MMC trial move in the canonical ensemble. Consequently, the present approach offers the advantage that NPs can easily adopt symmetric configurations with regular distributions of the elements, which would be hard to achieve with fixed compositions as the atomic ratios would have to match the \textit{a-priori} unknown equilibrium stoichiometries. The SGCE therefore removes the possible bias of predefined compositions and ensures that the equilibrium ratio of Cu and Zn atoms is reached independent of the initial state. In this way, for a given temperature and chemical potential difference, in the SGCE the composition is guided towards the optimum stoichiometry, allowing for an unbiased search of configurations with different compositions.

\section{Computational Details} \label{sec:comp_details}

\subsection{High-Dimensional Neural Network Potential Energy Surface}

In this work we use a HDNNP suitable for very large systems consisting of thousands of atoms that we have recently constructed for $\alpha$-brass NPs~\cite{JanPaper1}. 
It is based on reference DFT calculations of clusters, bulk and slab structures with varying ratios of Cu and Zn atoms within the $\alpha$-brass regime ($x_{\text{Zn}} < 40 ~\%$) employing the PBE\cite{Perdew1996a} exchange-correlation functional. The root mean squared errors (RMSE) of the total energies and forces for structures not included in the training of the HDNNP are $\SI{1.7}{\milli \electronvolt}$/atom and $\SI{39}{\milli \electronvolt \per \angstrom}$, respectively, ensuring a first-principles quality description. 
\\
The potential, which has been validated for a wide range of properties of large clusters, bulk $\alpha$-brass and its surfaces~\cite{JanPaper1}, can be used for different types of geometries, from the FCC lattice of $\alpha$-brass to the melt, while the periodic crystal structures that occur at higher Zn concentrations, which are not relevant for the present work. Instead we have explicitly included clusters with Zn concentrations higher than the $\alpha$-brass regime to ensure that the HDNNP will accurately describe the Zn surface accumulation in larger brass NPs\cite{JanPaper1}. We have thereby ensured that the HDNNP is applicable to situations like the accumulation of Zn atoms at the surface of NPs, which formally exceed the Zn-contents of $\alpha$-brass within the local atomic environments. Due to the underlying atomic environments used in the training process, in case of clusters the potential is applicable to systems containing more than approximately 75 atoms which is the smallest reference cluster included in the dataset. All details about the HDNNP construction and its validation can be found in our previous work \cite{JanPaper1}.

\subsection{Simulation Protocols} \label{sec:sim_setup}

\subsubsection{Metropolis Monte Carlo and Semi-Grand Canonical Ensemble} \label{sec:method_mmc}

All simulations have been performed with LAMMPS~\cite{Plimpton1995} including the n2p2 library for HDNNPs~\cite{Singraber2019}, which is compatible with the RuNNer program
~\cite{P4444,P5128} that has been used for the construction of the potential. 
The initial NP structures have been generated using Wulff constructions \cite{Wulff1901,ase-paper} of pure Cu clusters. The required surface energies of Cu have been determined by the HDNNP. Subsequently, a specified fraction of Cu atoms has been replaced by Zn atoms at randomly selected lattice sites\cite{JanPaper1}. 
\\
For an infinite number of MMC steps the final result of the simulation would only depend on the chosen chemical potential difference $\Delta \mu$ and the temperature $T$. To remove any bias in our results originating from the initial configurations we use different random initial compositions up to $x_{\text{Zn}} = 40 ~\%$, with various random element distributions and allow the systems to evolve starting from these points. 
\\
For our analysis, we define the average element-specific occupations for each individual site $i$ as
\\
\begin{align}
    \langle \rho_{\text{Cu}}^{i} \rangle = n_{\text{Cu}}^{i}/n
    \text{ and } 
    \langle \rho_{\text{Zn}}^{i} \rangle  = n_{\text{Zn}}^{i}/n ~,
    \label{eq:average_occ}
\end{align}
\\
where $n$ is the total number of configurations and $n_{\text{Cu}}^{i}$ and $n_{\text{Zn}}^{i}$ are the number of sampled configurations with Cu or Zn located at site $i$. These site-specific average occupations converge much slower than the global chemical compositions of the NPs.

\subsubsection{Simulated Annealing} \label{sec:method_sa}  \label{sec:method_md}

For the simulated annealing simulations reported in Sec.~\ref{sec:sim_ann}, we combined a series of $(NVT)$ molecular dynamics (MD) runs with a three-fold Nos\'{e}-Hoover chain thermostat\cite{Nose1984,P2756} and MMC element exchange moves at fixed composition. This allows for studying both variations in the NP shapes and elemental distributions. 

For each SA run, a random initial geometry has been generated by melting the NP at $\SI{1200}{\kelvin}$. Then, the final configuration from the melting trajectory has been used as initial geometry for the subsequent cooling protocol. First, the NPs have been cooled to $\SI{980}{\kelvin}$ which is just below the melting point of copper NPs at this size. As discussed below, the melting temperatures of brass nanoparticles are lower than of pure Cu NPs.
Next, a series of MD trajectories have been run starting from $\SI{980}{\kelvin}$ down to $\SI{250}{\kelvin}$ in intervals of $\SI{15}{\kelvin}$. 
At each temperature a MD simulation has been performed with a duration of 112~ps and a time step of $\Delta t = \SI{1.4}{\femto \second}$ resulting in a total simulation time of about $\SI{5.6}{\nano \second}$ per NP. After each individual MD simulation, the element distribution of the final structure has been partially optimized by employing 400 MMC Cu/Zn exchange moves at the temperature of the respective MD simulation. The goal of these MMC steps is to increase the efficiency of the redistribution of the elements, which would be very slow particularly at the lower temperatures in conventional MD simulations, without aiming for fully optimized configurations. This alternation between MD and MMC has been continued until finally the target temperature of $T=\SI{250}{\kelvin}$ has been reached. At this final temperature, the distribution of elements within the NP has been optimized using a larger number of $20000$ MMC steps. In the results section, we will refer to this protocol as SA1. Furthermore, a second protocol SA2 has been established using smaller temperature intervals of $\SI{10}{\kelvin}$, a final temperature of $T=\SI{200}{\kelvin}$, and a much larger number of $1000000$ MMC steps at the end of the simulation to check on the convergence of the determined properties.

For the combined MMC and MD simulations discussed in Sec.~\ref{sec:dynamic_lattice} the NPs have been heated starting from $\SI{150}{\kelvin}$ up to $\SI{1400}{\kelvin}$ using steps of $\SI{10}{\kelvin}$. For each temperature increment we have performed an MD simulation of $\SI{0.28}{\nano \second}$. Every 200 MD time steps we performed 100 MMC exchange moves in the SGCE to efficiently search for energetically favorable distributions of both elements. Here we set the chemical potential to a fixed value of $\Delta \mu = \SI{-2.4}{\electronvolt}$ which provides a stable balance of Cu to Zn for most temperatures and NP sizes.

\subsection{Structural Analysis}

Several lattice order and structure parameters as implemented in the OVITO\cite{P5515} molecular visualization program have been used to analyse the results of the simulated annealing runs.
\\
The centrosymmetry parameter\cite{centrosymm} is defined as a sum over the nearest neighbours of a central atom,
\begin{align}
    p_{\text{CSP}} = \sum_{i=1}^{N/2} | \Vec{r}_i  + \Vec{r}_{i+N/2}|^2~,
\end{align}
where $\Vec{r}_i$ and $\Vec{r}_{i+N/2}$ are vectors pointing to the two opposite neighbours of the central atom and $N=12$ for FCC, hexagonal closed-packed (HCP) and $N=8$ for body-centered cubic (BCC), respectively. In a non-symmetric neighbourhood with crystal defects or surfaces $p_{\text{CSP}}$ will take a large positive value. On the other hand it will be zero for atoms inside a perfect crystal. 
The centrosymmetry parameter is used in solid-state systems to measure the local disorder around a central atom and it is useful to decide whether an atom is part of an ordered lattice with local homogeneous values $p_{\text{CSP}}$.
Note that for visualization purposes we will normalize $p_{\text{CSP}}$ by the largest atomic value of $p_{\text{CSP}}$ within each crystal structure.
\\
Polyhedral Template Matching\cite{P5714} (PHTM) serves as a robust classifier for the local crystal structure and is particularly useful if the lattice is subject to strain when methods like common neighbour analysis\cite{cna} cannot be applied reliably. This is used to determine which crystal structure lattice (HCP, FCC, etc.) the local environment belongs to.

\section{Results}\label{sec:results}

In the following Sections we discuss NP structures obtained from various types of simulations using energies and forces provided by the HDNNP. First, we study the Cu/Zn distributions on a fixed FCC lattice representing the atomic positions in the NPs. Subsequently, the role of lattice relaxation is investigated showing that the obtained results also hold if lattice strain is removed (s. SI Sec.~1.2.2). Finally, as simple lattice relaxations are not expected to allow escaping the structural models defined by the initial Wulff shapes of the NPs, we use SA simulations to examine global structural changes.

\subsection{Simulations in the Semi-Grand Canonical Ensemble}  \label{sec:results_sgce}

\subsubsection{Nanoparticles with Fixed Lattice}

\begin{figure}[h!]
         \centering
          \includegraphics[width=\columnwidth]{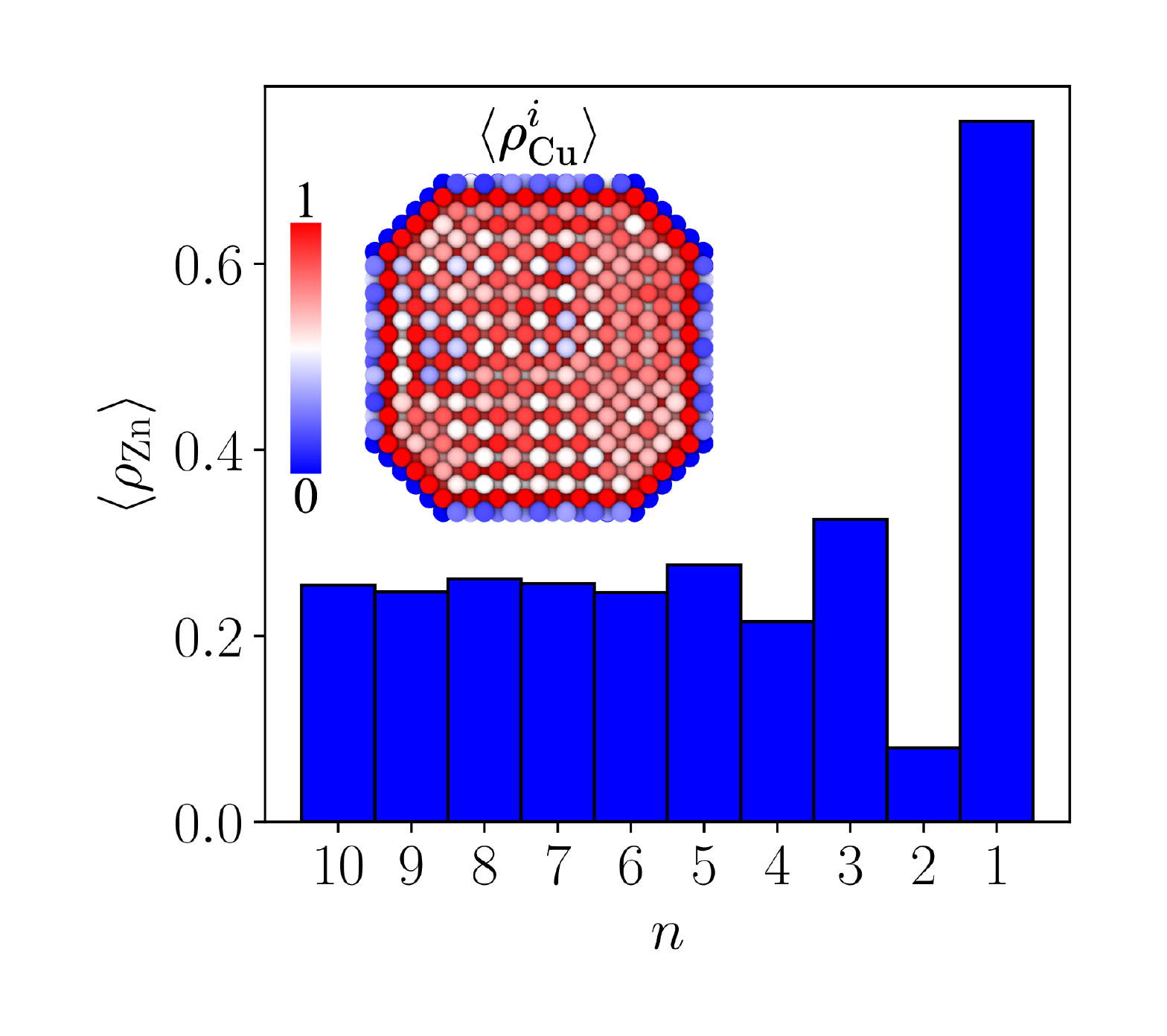}
         \caption{
         Average Zn occupation $\langle \rho_{\text{Zn}} \rangle$ at $T = \SI{200}{\kelvin}$ and $\Delta \mu = -\SI{2.55}{\electronvolt}$ within each shell of atoms for a brass nanoparticle consisting of 4897 atoms and a radius of $\approx \SI{25}{\angstrom}$, where $n=1$ is the outermost shell. In the inset we show for a cross section of the cluster the site-resolved Cu occupation $\langle \rho^i_{\rm Cu} \rangle=1-\langle \rho
        ^i_{\text{Zn}} \rangle$. The average occupation of a lattice site is given by the color (red: $100~ \%$ occupied by Cu, blue: $100~ \%$ occupied by Zn, white: equal occupations). All atomic visualizations in this work have been generated with the software OVITO~\cite{P5515}. 
            }
    \label{fig:central_dist} 
\end{figure}

As a first step we performed SGCE Monte Carlo simulations of Wulff-shaped NPs with fixed lattices to determine values of $\Delta \mu$ resulting in stable mixtures of Cu and Zn in the NPs. For instance, a NP containing 4897 atoms with $\Delta \mu = \SI{-2.55}{\electronvolt}$ at $T=\SI{200}{\kelvin}$ leads to a Zn atom fraction of about $30 ~\%$ (Fig.~\ref{fig:central_dist}). An interesting general observation is that the outermost shell and in particular low coordination sites such as edges and corners exhibit the highest Zn atom fraction. On the other hand, the Zn atom fractions of the sites in the interior of the NP are much lower, while the smallest Zn atom fraction is found in the first sub-surface layer, which is almost exclusively occupied by copper atoms.

\begin{figure}[htb]
          \centering           
          \includegraphics[width=\columnwidth]{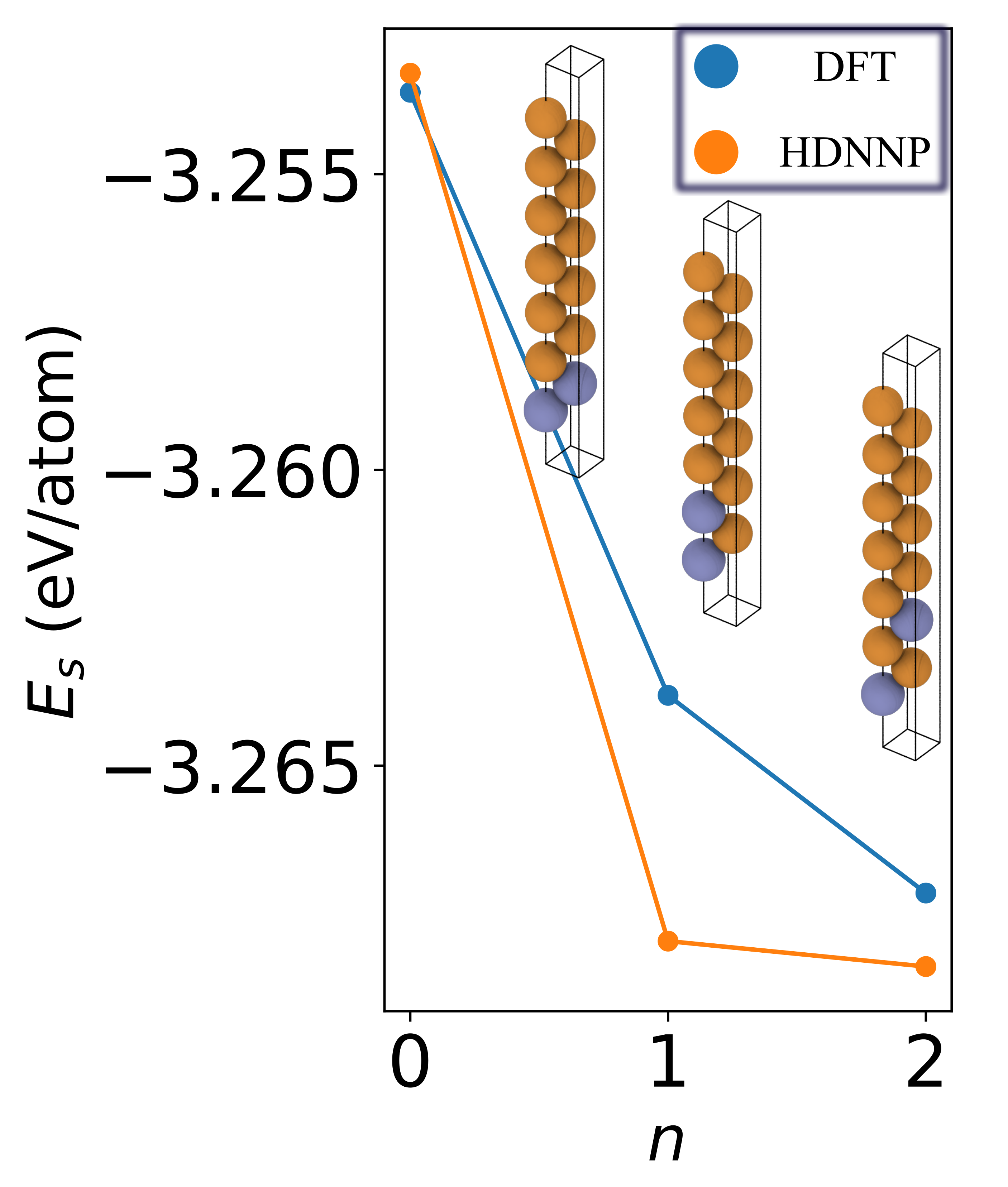}
          \caption{DFT and HDNNP energies of relaxed copper slabs with a double layer of Zn. One layer of Zn is fixed at the surface while the second is moved towards the center as illustrated by the three slabs.
          }
     \label{fig:slab_configurations} 
 \end{figure}

Notably we did not observe formation of a Zn double layer. To illustrate this specific arrangement of Cu and Zn layers close to or at the surface we computed the energy for fully relaxed Cu(100) slabs with a second zinc layer as a function of the distance from the surface while the first zinc layer remains in the topmost layer (s. Fig.~\ref{fig:slab_configurations}). We find that the configuration with two Zn layers stacked next to each other corresponds to the highest energy as predicted by the HDNNP and also DFT. Therefore, accumulation of copper in the second layer is indeed significantly more beneficial than accumulation of Zn which explains why a double Zn layer was never observed in a simulation.

\begin{figure}[h!]
         \centering
          \includegraphics[width=\columnwidth]{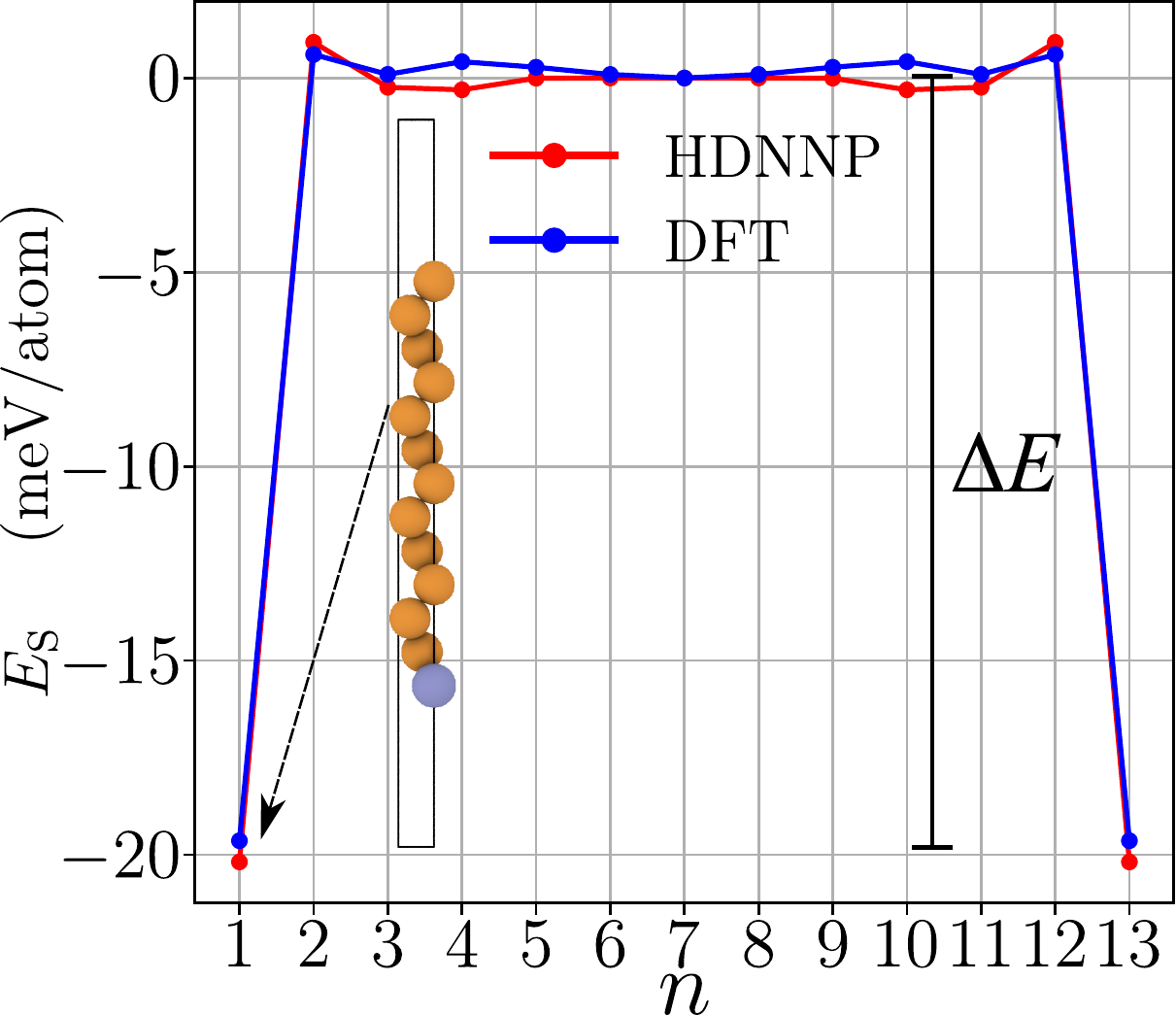}
         \caption{
         Relative energies $E_{\text{S}}$ of 13-layer $(1 \times 1)$ Cu(111) slabs containing a single Zn monolayer in layer number $n$ relaxed with DFT (red) and the HDNNP (blue). Energy scale is shifted such that a slab with Zn in the middle ($n=7$) defines the zero energy point for each method. The first ($n=1$) and the last ($n=13$) point correspond to a Zn surface layer.
         }
    \label{fig:brass_slab} 
\end{figure}

To understand the surface accumulation of Zn atoms we performed MMC simulations of periodic brass slabs representing the (111), (110) and (100) surfaces. Also for the top layers in these systems we observed an accumulation of Zn atoms forming regular surface patterns (s. SI Fig.~14) and Zn depletion in the first subsurface layer. 
\\
To further analyze this phenomenon, we calculated DFT energies of fully relaxed 13-layer Cu(111) slabs containing a single monolayer of Zn atoms at different positions in the slab (Fig.~\ref{fig:brass_slab}) using the VASP code~\cite{P0156,Kresse1999} with the same setup that has been used in the reference calculations~\cite{JanPaper1}. These calculations confirmed that Zn in the outer-most layer is the energetically most favorable configuration, while the incorporation of Zn in the first subsurface layer results in the highest energy.
The DFT energy difference of the relaxed Cu(111) slabs with the Zn layer in the center or at the surface, respectively, shown as $\Delta E$ in Fig.~\ref{fig:brass_slab}, is $\Delta E^{\rm DFT} = \SI{19.6}{\milli \electronvolt}$ per atom, which is very close to the value of $\SI{20.2}{\milli \electronvolt}$ per atom predicted by the HDNNP for this system. 
We have repeated these calculations also for the Cu(110) and Cu(100) surfaces with very similar results (s. SI Fig.~11). 
Hence, Zn accumulation at the surface is neither an artifact of the fixed lattice that might induce strain and thus an energy increase of the system upon incorporation of Zn atoms nor of the HDNNP. 
\\
In agreement with our results, migration of Zn atoms to brass surfaces has previously been observed in experiment \cite{Sano2002} and there may be two possible explanations. First, the cohesive energy of pure Cu is significantly larger than the cohesive energy of Zn, with Zn-Cu interactions exhibiting an intermediate strength. Thus, it is energetically favorable to minimize the interactions involving Zn atoms, by segregating them to the under coordinated surface sites \cite{reviewerref, C9CP06967D}. Second, Zn atoms have a larger covalent radius than Cu atoms, as the nearest neighbor distance in metallic HCP zinc \cite{zinc_lattice} is $\SI{2.66}{\angstrom}$, compared to $\SI{2.55}{\angstrom}$ in FCC Cu~\cite{Davey1925}, facilitating Zn surface accommodation to minimize the distortion to the rest of the lattice. 
\\
The accumulation of Zn atoms at the surface suggests that the favored composition may depend on the NP size, since the surface to bulk ratio decreases as the NP size increases. To test this hypothesis we have determined the composition of NPs containing $79, 459$, and $1103$ atoms as a function of the chemical potential difference $\Delta \mu$ at $T=\SI{200}{\kelvin}$ allowing to search for low energy structures while keeping a reasonable acceptance ratio in the MMC simulations.
In accordance with the definition of $\Delta \mu$, the Zn atom fraction $\langle x_{\text{Zn}}\rangle$ increases as $\Delta \mu$ becomes more negative, and several interesting observations can be made.
\\

\textbf{Size and Composition}:
\\

The relation between composition and NP size in Fig.~\ref{fig:isoNomin} reveals that $\Delta \mu$ must be more negative for larger NPs to reach the same Zn atom fraction $x_{\text{Zn}}$ compared to smaller NPs. The extreme case is represented by bulk brass (s. SI Fig.~2) which could be considered an infinitely large NP without surface. 
This is because the ratio of surface-to-bulk sites decreases with increasing system size such that the relative abundance of energetically favorable surface sites for Zn atoms decreases, preventing the accommodation of a large number of Zn atoms at the surface. 
\\
Instead, in larger NPs Zn atoms must be incorporated into the interior of the system, which is energetically unfavorable in form of an increased cohesive energy that must be compensated by a higher chemical potential of zinc or a lower chemical potential difference $\Delta \mu$, respectively. The underlying increase of the cohesive energy of the CuZn system with $x_{\text{Zn}}$ has been described in detail in our previous work~\cite{JanPaper1}. 
\\

\begin{figure}[htb]
          \centering           \includegraphics[width=\linewidth]{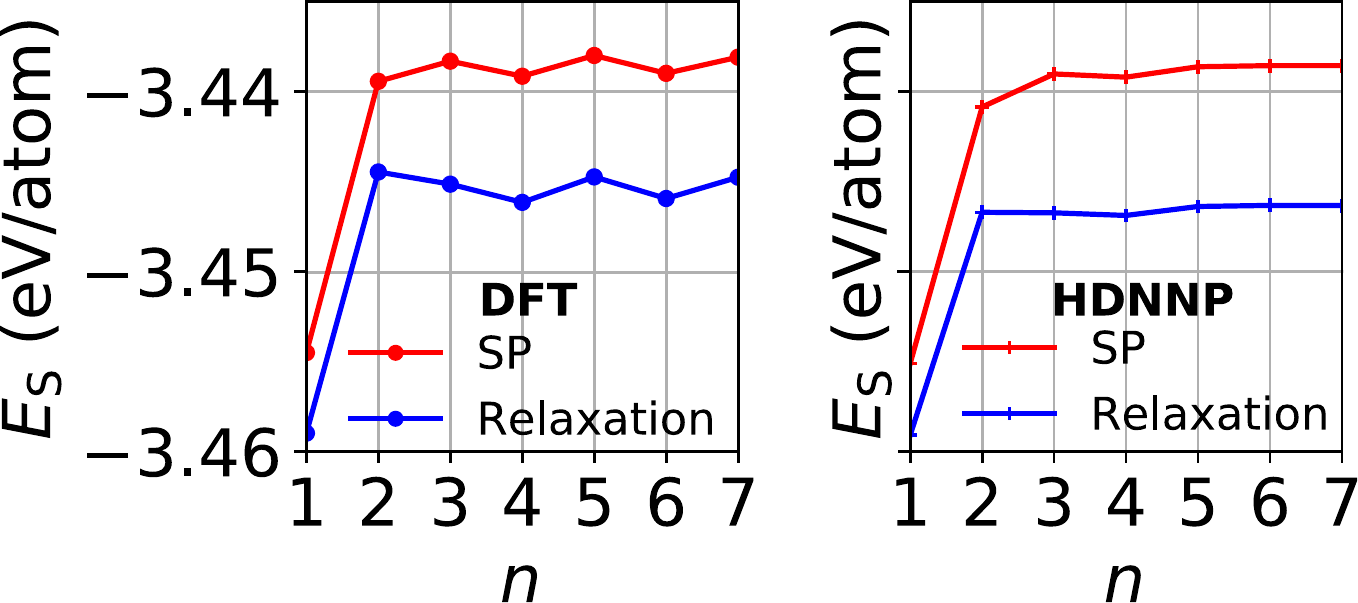}
          \caption{Comparison of DFT energies $E_s$ of a Cu (100) slab with a single layer of Zn at position $n$ resulting from single point calculations (SP) and relaxation. Here $n=1$ corresponds to Zn exposed to vacuum and $n=7$ is the center of the slab (with $n=13$ layers in total).
          }
     \label{fig:relaxation_slabs} 
 \end{figure}

\begin{figure}[h!]
         \centering
          \includegraphics[width=\columnwidth]{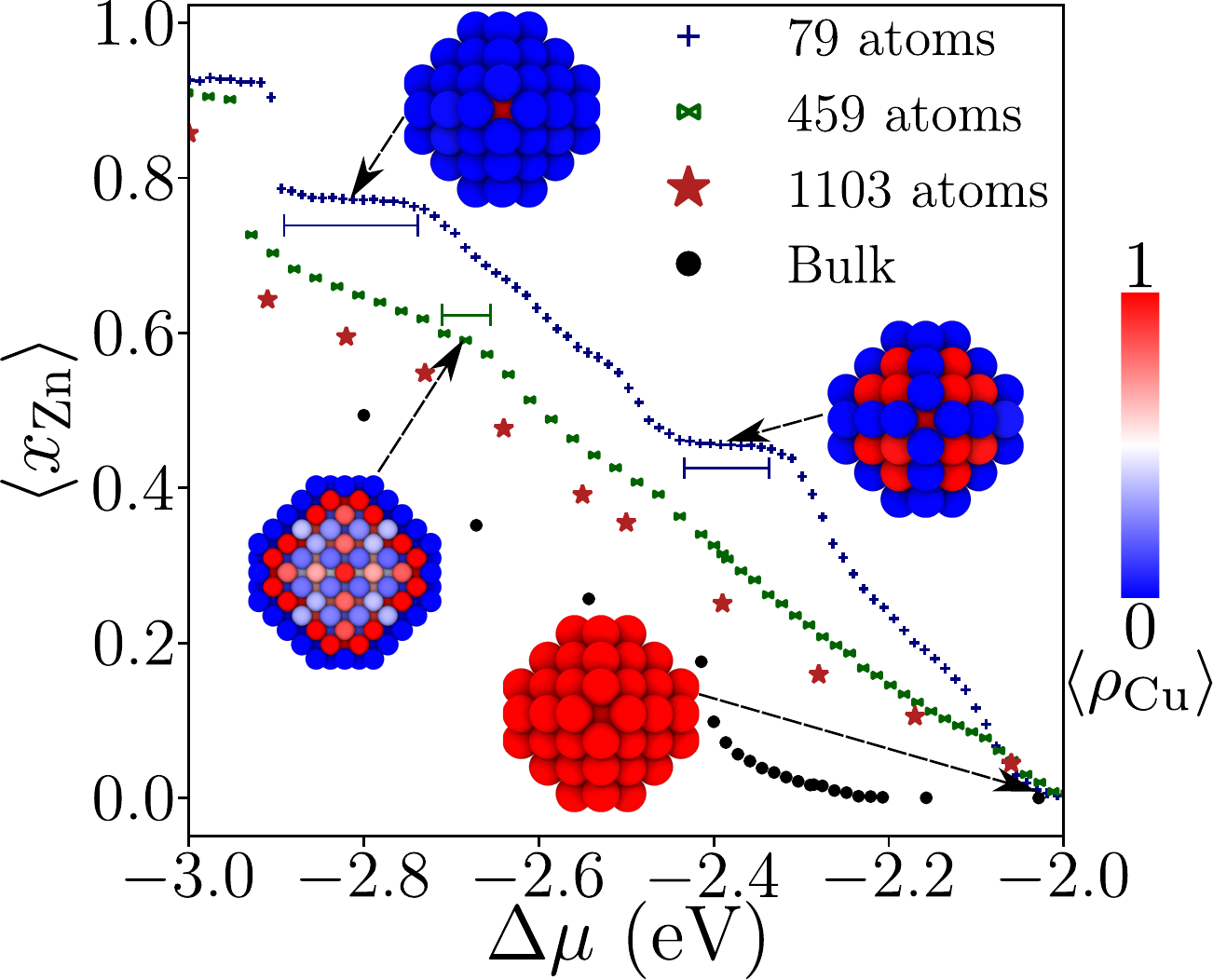}
         \caption{
         Average Zn atom fraction $\langle x_{\text{Zn}} \rangle$ as a function of chemical potential difference $\Delta \mu$ at $T=\SI{200}{\kelvin}$ for NPs containing 79, 459 and 1103 atoms and for
         a $4\times 4\times 4$ bulk supercell containing 256 atoms. Plateaus indicate $\Delta \mu$ intervals corresponding to NPs with a particularly stable structure. Red atomic sites label a predominant occupation by Cu, while blue represents Zn. 
                }
    \label{fig:isoNomin} 
\end{figure}

Note that because of the strain in the system due to the fixed lattice and the larger Zn nearest neighbor distance, the energies of NPs with high Zn contents will be overestimated.
To expand on this observation we have performed DFT calculations of the energy of Cu (100) slab with a single Zn layer, both with and without relaxation and we compare the resulting energy of the slab for both the HDNNP and DFT (s. Fig.~\ref{fig:relaxation_slabs}). 
Note that here we did not subtract the energy of the configuration with Zn in the center, because a comparison of different energy scales of the same system is needed to understand the energy difference due to relaxation. Based on DFT as well as HDNNP calculations we find that the relaxed structures are about $5 ~\text{meV/atom}$ more stable than their corresponding bulk-terminated slab.

Probably the most significant observation from these relaxations of the Cu (100) slab with a single Zn layer is that the configuration with Zn exposed to vacuum is substantially ($\SI{15}{\milli \electronvolt \electronvolt}$/atom according to DFT/HDNNP) more stable than the configuration with Zn in the center of the slab.

Thus to obtain more realistic energies, we have performed an additional series of simulations where the atomic positions are also relaxed. 

As expected this generally results in higher Zn atom fractions in larger NPs for the same chemical potential as in the unrelaxed configuration. However, the overall trend that Zn atoms are more difficult to incorporate in larger NPs essentially remains unchanged.
These results for the composition and next neighbour distances of nanoparticles when including lattice relaxations are discussed in more detail in the SI in Sec.~1.2.2.
\\

\textbf{Concentration Plateaus}: 
\\

Further analyzing Fig.~\ref{fig:isoNomin} we find that in particular $\langle x_{\text{Zn}} \rangle $ of the smallest cluster containing only 79 atoms exhibits plateaus where the average composition remains almost constant in spite of notable changes in $\Delta \mu$. The two most prominent plateaus correspond to patterns with Zn-covered edges and corners ($\SI{-2.4}{\electronvolt}<\Delta \mu<-\SI{2.3}{\electronvolt}$) or a completely Zn-covered surface ($\SI{-2.9}{\electronvolt}<\Delta \mu<-\SI{2.8}{\electronvolt}$), respectively. 
Another plateau can be found for the 79 and 459 atom NPs at $\Delta \mu<-\SI{2.9}{\electronvolt}$ where for the 79 atom NP all lattice sites except for a single central atom are occupied by Zn and for the 459 atom NP most sites inside the NP are occupied by Zn.

These structures correspond to compositions of high stability in terms of a low potential energy. Similar plateaus corresponding to complete outermost Zn shells at more positive $\Delta \mu$ values could not be found for the larger $459$ or $1103$ atom NPs, but it can  be observed that the Zn concentration curve is slightly flattened in certain $\Delta \mu$-intervals. The absence of distinct plateaus in the larger clusters is a result of the broader range of potential energy changes when substituting Cu by Zn atoms in these systems compared to the small cluster with a much smaller diversity in atomic sites.
\\

\textbf{Zinc Atom Patterns}: 
\\
\begin{figure*}[h!]
         \centering
         \includegraphics[width=\linewidth]{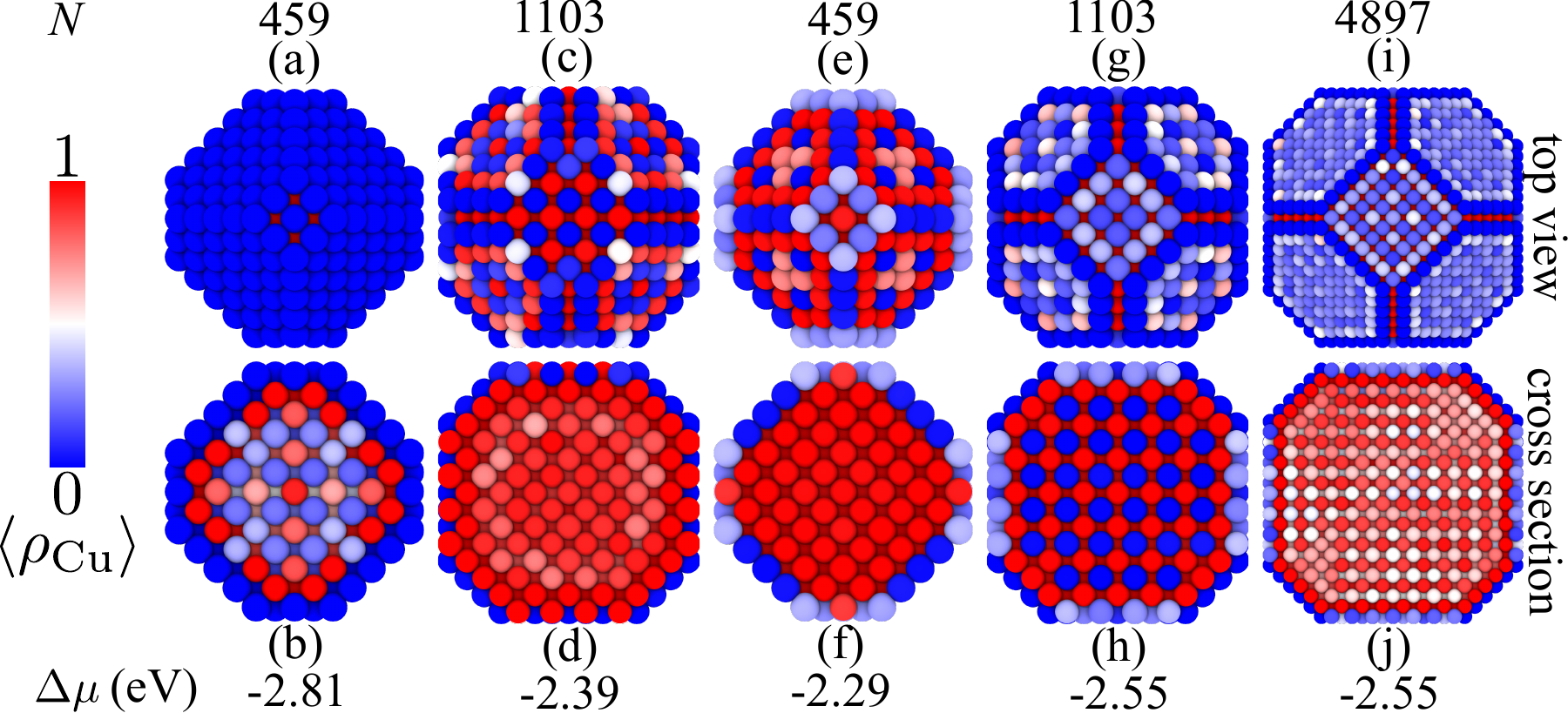}
         \caption{
         Average Cu occupation $\langle \rho_{\text{Cu}} \rangle$ at $T= \SI{200}{\kelvin}$ for fixed lattice NPs of different sizes and chemical potential differences $\Delta \mu$. The upper row shows top views and the lower row the corresponding cross sections. Note that the average occupations for the 4897 atom NP shown in panels (i) and (j) are not fully symmetric as the simulations for systems of this size are difficult to converge.
         }
    \label{fig:all_long_sim} 
\end{figure*}

Interestingly, for specific $\Delta \mu$ values very symmetric average elemental site occupations can give rise to characteristic structural motifs of Zn and Cu (Fig.~\ref{fig:all_long_sim}), e.g. core-shell configurations or regular surface patterns in case of the NPs consisting of $459$ and $1103$ atoms, respectively. 
Such Zn triangles or hexagons on (111) (Fig.~\ref{fig:all_long_sim}e,c) or squares on (100) surfaces (Fig.~\ref{fig:all_long_sim}c,g) were also found for brass surfaces (SI Fig.~14). Again we find that the outermost shell of the NPs, e.g. in Fig.~\ref{fig:all_long_sim}a, g and i, is predominantly or even completely occupied by Zn atoms while the first subsurface shell of lattice sites exhibits a very high Cu fraction (Fig.~\ref{fig:all_long_sim}b,h). If the system is forced to accommodate a high Zn concentration by a very negative $\Delta \mu$ value, usually an outer Zn shell is formed along with a heterogeneous core (Fig.~\ref{fig:all_long_sim}a,b). Formation of two outermost shells both predominantly occupied by Zn or even separated Zn and Cu domains have not been observed, which is a consequence of the more negative cohesive energy of brass.

\begin{figure*}[h!]
    \centering
    \includegraphics[width=0.7\linewidth]{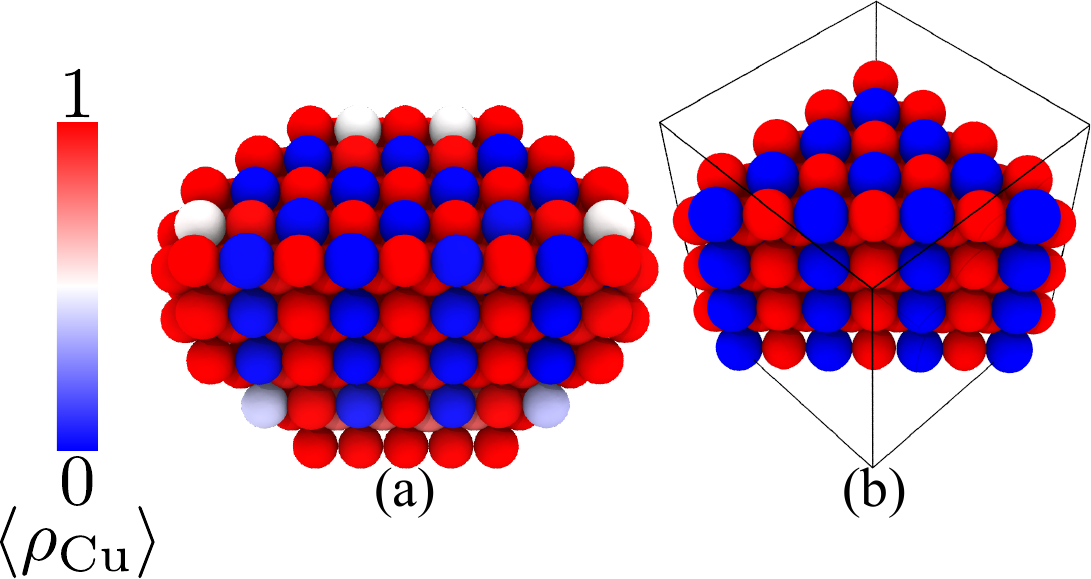}
    \caption{
    Cross section of the 1103 atom brass NP shown in Fig.~\ref{fig:all_long_sim}g (a). The interior occupation density $\langle \rho_{\text{Cu}} \rangle$ reveals the formation of the $L_{12}$ $\alpha$-brass low-energy phase~\cite{Muller,JanPaper1} of composition $\text{Cu}_{0.75}\text{Zn}_{0.25}$ ($\Delta \mu = \SI{-2.55}{\electronvolt}$, $T=\SI{200}{\kelvin}$). For comparison, (b) shows a cut of a 256 atom $L_{12}$ bulk cell of the ideal $L_{12}$ $\alpha$-brass phase. }
    \label{fig:l12nano_bulk}
\end{figure*}

For the 459 atom NP shown in Figs.~\ref{fig:all_long_sim}b and f it is found that first the outermost shell is fully occupied by Zn before additional Zn atoms are incorporated in the interior for a more negative $\Delta \mu$. For larger NPs containing 1103 atoms Zn can also be found in the center before the outermost shell is completely occupied by Zn (Fig.~\ref{fig:all_long_sim}d and h). In this case the reason is the  relatively high stability of an ordered bulk phase in the interior of the NP (Fig.~\ref{fig:all_long_sim}h). Specifically, at $\Delta \mu = \SI{-2.55}{\electronvolt}$ we find the  $L_{12}$ phase\cite{Muller} of $\alpha$-brass with composition $\text{Cu}_{0.75}\text{Zn}_{0.25}$ in the interior region of the NP. For comparison, the corresponding structure of the periodic bulk phase is shown in Fig.~\ref{fig:l12nano_bulk}. Several low-energy bulk phases  of $\alpha$-brass have been proposed in the literature, like the $L_{12}$, $LPS_3$, $DO_{23}$ phases \cite{P2119}, which could also be found in MMC simulations of a 256 atom supercell employing the HDNNP in our earlier work~\cite{JanPaper1}. It could be speculated that with a much increased number of MMC steps, a low-energy bulk phase could also be obtained for the 4896 atom NP since the required Zn concentration of $25 ~\%$ is present in the center of this NP (s. Fig.~\ref{fig:central_dist}). The formation of a low-energy bulk phase in combination with Zn enrichment at the surface in general enables the systems to adopt a much more stable configuration. For very large systems, the number of bulk lattice sites increases substantially and will finally dominate over the surface sites. 

It is possible that for very large nanoparticles where low energy bulk configurations dominate corresponding plateaus in the zinc atom fraction $x_{\text{Zn}}$ as a function of chemical potential will be observed.

In summary, depending on the NP size a certain amount of Zn atoms may accumulate in the interior region as a result of a subtle interplay between energy and proportion of available sites. As a consequence, sites at the surface can be inhibited from reaching full Zn occupation as shown for example in Figs.~\ref{fig:all_long_sim}g and i. In these cases the surface sites are only partially occupied by Zn atoms, and instead a regular Zn pattern is emerging in the core region (Figs.~\ref{fig:all_long_sim}h and j). Consequently, an equilibrium is established between the surface and bulk interior of the NP, competing for the Zn atoms available at a given chemical potential.

\begin{figure}
\centering
         \includegraphics[width=\columnwidth]{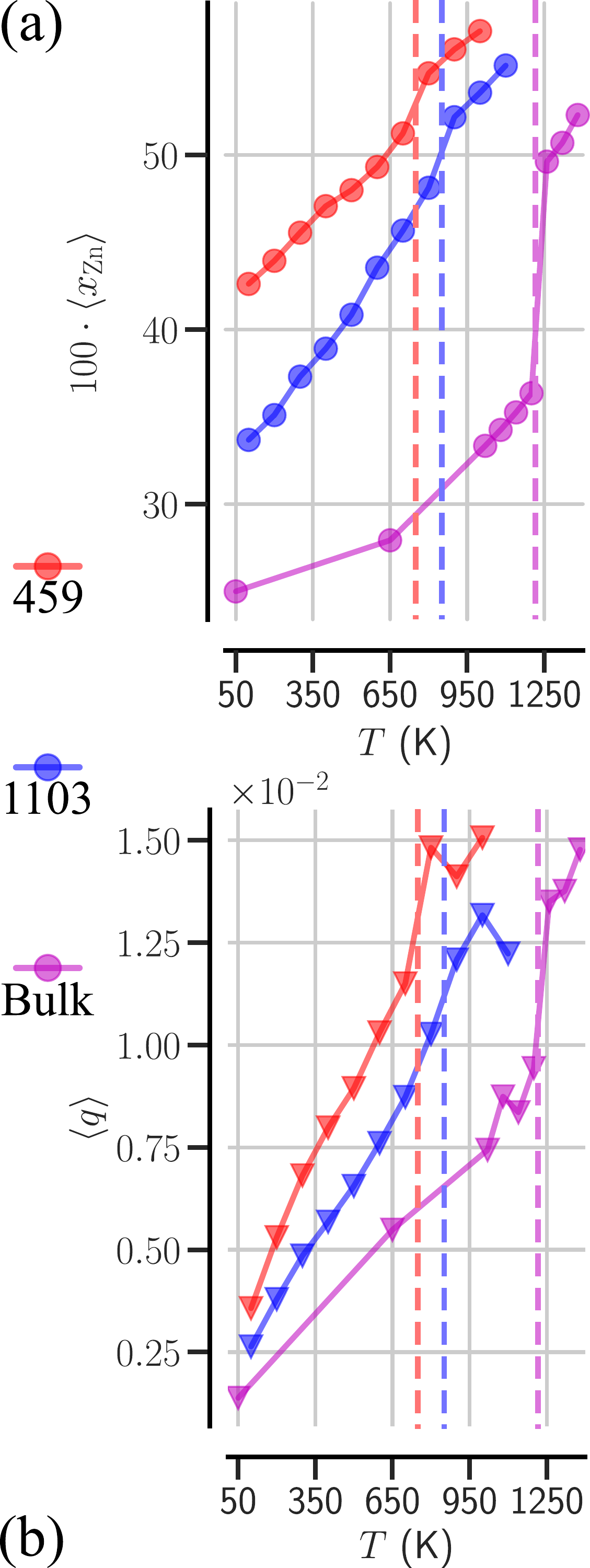}
         \caption{
         Average Zn atom fraction $\langle x_{\text{Zn}} \rangle$ (a) and Lindemann index\cite{Lindemann1910} values $\langle q \rangle$ (b)
         for brass NPs containing $459$ and $1103$ atoms and a 256 atom bulk supercell as a function of $T$ at fixed $\Delta \mu = \SI{-2.4}{\electronvolt}$. The vertical dashed lines show the respective melting points.
         }
         \label{fig:dynamics_sgce}
\end{figure}

\subsubsection{Nanoparticles with Flexible Lattice}  \label{sec:dynamic_lattice}

In the previous Section we have seen that even with a fixed lattice often qualitative insights into the structure of brass NPs can be obtained. Nevertheless, in the next step we will now abandon the approximation of a fixed atomic lattice by allowing the atoms to relax. The resulting lowering of the potential energy is expected to slightly increase the acceptance probability in particular for structures with higher Zn fractions $x_\text{Zn}$ in MMC SGCE simulations. We will now study this effect in more detail with hybrid MD/Monte Carlo simulations, in which MMC exchange moves in the SGCE allow sampling the composition as a function of $T$.

Fig.~\ref{fig:dynamics_sgce} shows the Zn atom fractions $\langle x_{\rm Zn} \rangle$ and Lindemann indices~\cite{Lindemann1910, JanPaper1} $\langle q \rangle$, serving here as a measure for the average deviation of the atomic positions from their equilibrium positions, as a function of $T$ for NPs containing 459 and 1103 atoms as well as for a 256 atom bulk system for $\Delta = \SI{-2.4}{\electronvolt}$. For each system, $\langle x_{\rm Zn} \rangle$ and $\langle q \rangle$, show a very similar behavior and for a given system exhibit discontinuities at about the same temperature corresponding to the melting point $T_m$~\cite{JanPaper1}.
As temperature is increased potential energy differences will become less dominant for the MMC and the acceptance of zinc-rich structures increases.
\\
The estimated melting point of the bulk brass phase which can be extracted from the sharp transition (s. Fig.~\ref{fig:dynamics_sgce}) at about $T_{m} = \SI[separate-uncertainty = true]{1200(10)}{\kelvin}$ with a Zn atom fraction of $34~\%$ is surprisingly close to the experimental value of $T_{m}^{\text{exp}}  = \SI{1203}{\kelvin}$ at $x_{\text{Zn}} = 33 \%$\cite{brass}. 
While deriving an estimated value of $T_{m}$ for the NPs directly from the Lindemann index\cite{Lindemann1910} (s. Fig.~\ref{fig:dynamics_sgce}) is possible, we follow a more systematic approach and extract $T_{m}$ from a parameter corresponding to the turning point of curves fitted to the Lindemann index values as described in [\citenum{JanPaper1}].
This results in $\SI{736}{\kelvin}$ for the 459 atom and $\SI{799}{\kelvin}$ for the 1103 atom NP respectively which we also marked as vertical dashed lines in Fig.~\ref{fig:dynamics_sgce}.

Comparing the Zn atom fractions of the 459 atom NP for a low temperature of $T \approx \SI{100}{\kelvin}$ and after melting we find that the Zn atom fraction is increased by $10~\%$ at elevated temperatures, while it increases by about $15~\%$ for the 1103 atom NP (s. Fig.~\ref{fig:dynamics_sgce}).
\\
An increase in the Zn concentration with temperature was also found for a static atomic lattice albeit only between $3$-$5~\%$ (s. SI Fig.~13a Sec.~1.2.1 in SI), which is in accordance with the necessity that the occupation of sites converges to a ratio of 1:1 for Zn and Cu at infinite $T$. This is true for any two state system, and is more pronounced even for lower $T$ the closer the energies of the two states are.

\subsection{Simulated Annealing}\label{sec:sim_ann}

\begin{figure*}
         \centering
          \includegraphics[width=15cm]{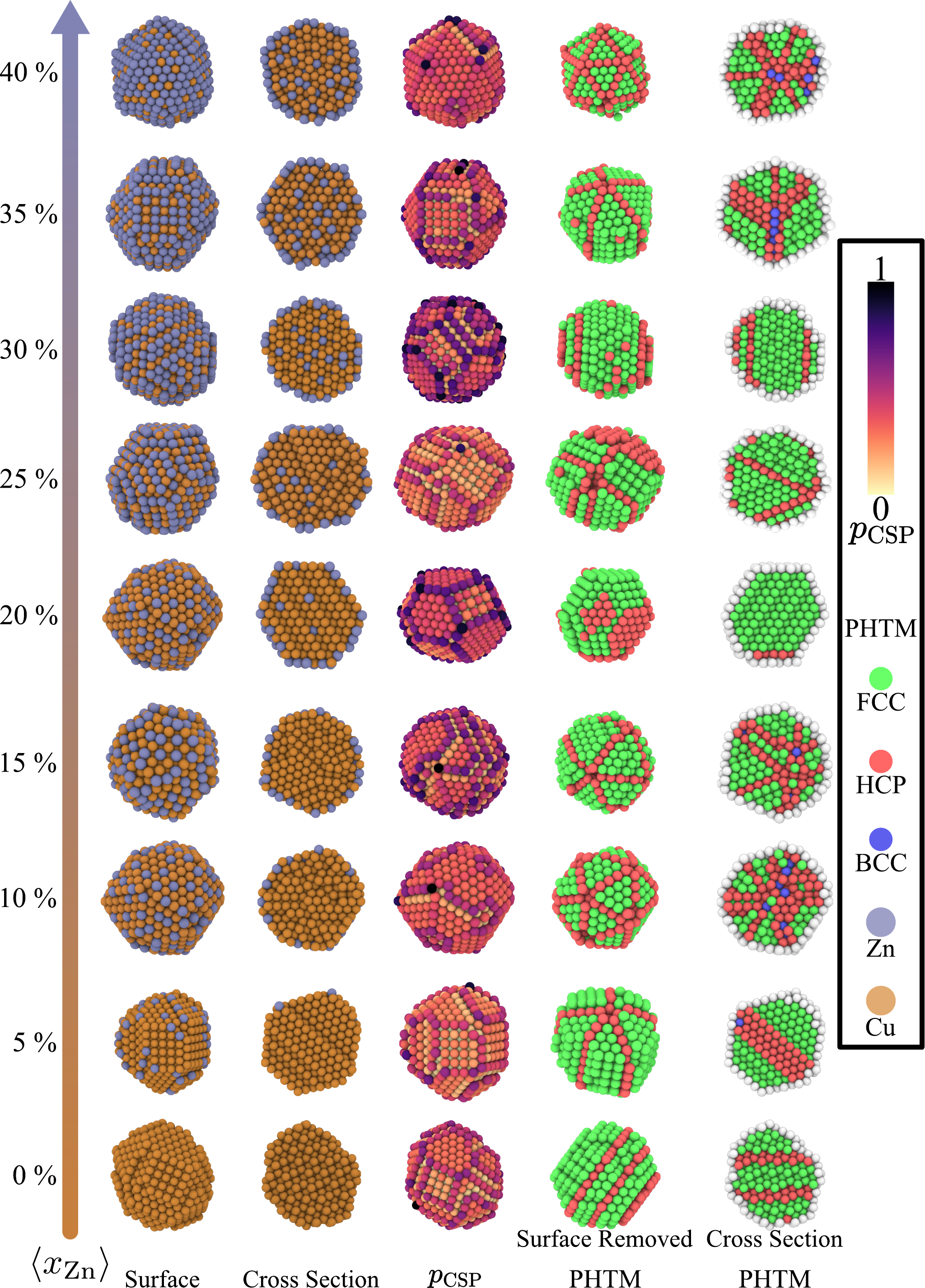}
 \caption{
  Brass NPs containing 1103 atoms obtained in simulated annealing according to protocol SA1 (diameter $\approx \SI{3}{\nano \meter}$) for different Zn atom fractions. Shown are from left to right the element distribution at the surface, a cross section, the centrosymmetry parameter \cite{centrosymm}, and a classification of the atomic environments using the Polyhedral Template Matching\cite{P5714,P5515} (PHTM). For PHTM the atoms of the outermost layer are not shown, green atoms have a local FCC and red atoms a local HCP environment.
 }
    \label{fig:sim_ann}
\end{figure*}

The SGCE MMC simulations in the previous section have all been started from initial structures obtained from a Wulff-construction of pure copper NPs. Although lattice relaxation has been considered, it is very unlikely that major structural changes will occur in these simulations. Therefore, we will now investigate the possible existence of other low-energy geometries by SA.  
First, we used the SA1 protocol described in Sec.~\ref{sec:method_sa} to study brass NPs containing 1103 atoms for different compositions (Fig.~\ref{fig:sim_ann}). We mainly observe the formation of (111) and (100) surfaces for both low and high Zn atom fractions, with the (111) surfaces covering a higher fraction of the surface area than the (100) surfaces. Consequently, well-ordered clusters are formed from the melt, which are not too different from those obtained in a Wulff-construction. As observed before, Zn atoms are primarily found in the outermost layer.

\begin{figure}
         \centering
          \includegraphics[width=0.35\textwidth]{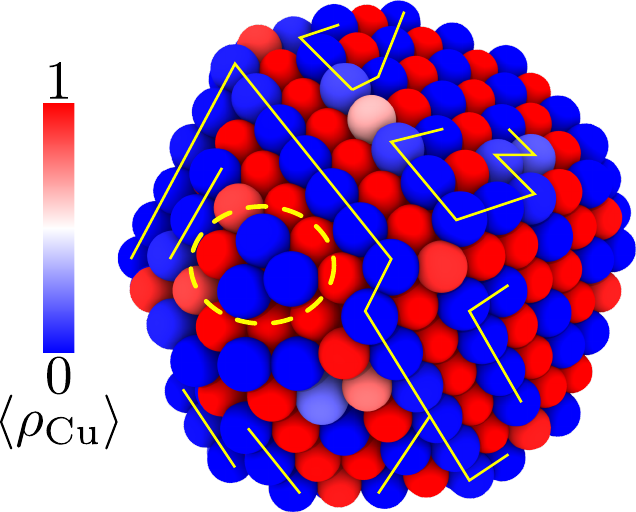}
 \caption{
 Site occupation by copper atoms $\langle \rho_{\text{Cu}} \rangle$ in a 1103 atom brass NP (using protocol SA2) for a Zn atom fraction of $x_{\text{Zn}}=20~\%$. Rows of Zn atoms are marked by yellow lines. A ``Zn-island'' is encircled with a dashed yellow line. For the MMC sampling of the occupations the final structure of the simulation has been used.
 }
    \label{fig:rows}
\end{figure}

Interestingly, we find rather well-ordered patterns of the Zn atoms at the surface resulting in the formation of regularly distributed isolated Zn atoms for low Zn contents or rows of Zn atoms for higher contents, as can be seen e.g. in Fig.~\ref{fig:rows} for a Zn fraction of $20~\%$, obtained with the SA2 protocol. The formation of rows is likely to be a result of strain minimization at the surface due to the larger Zn atoms. Cross sections of the NPs (Fig.~\ref{fig:sim_ann}) show that no significant amount of Zn is found in the center for $x_{\text{Zn}}$ smaller than $20~\%$.
\\

\begin{figure*}
         \centering
          \includegraphics[width=\linewidth]{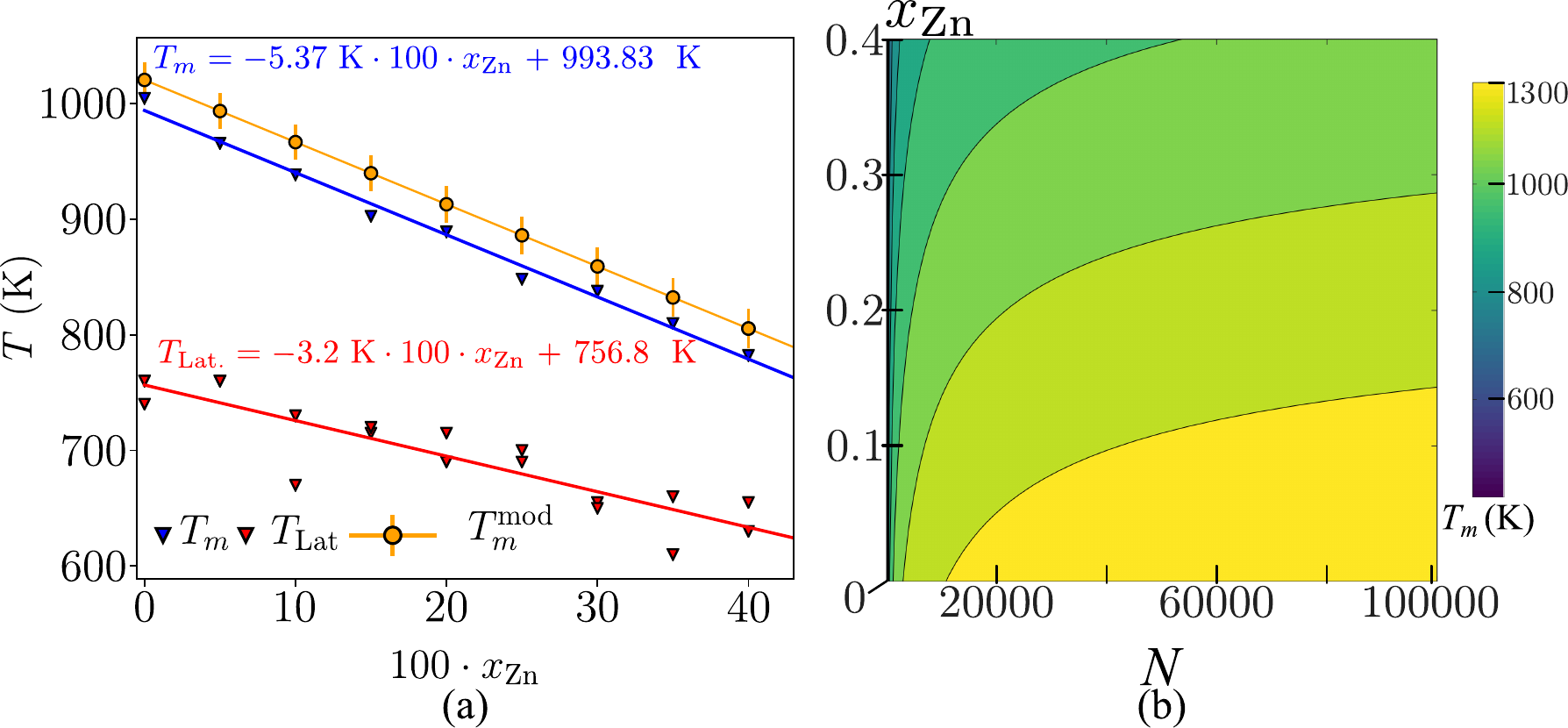}
 \caption{ 
 Melting temperature $T_m$ (blue) of brass NPs containing 1103 atoms obtained from an analysis of the Lindemann index, $T_{m}^{\text{mod}}$ predicted by Eq.~(\ref{eq:melting_curve}) (orange) and lattice formation temperatures $T_{\text{Lat}}$ (red) with two corresponding linear fits as a function of Zn content $x_{\text{Zn}}$ (a). In (b) we show a surface plot of the relationship between $T_m$, $x_{\text{Zn}}$ and the number of atoms $N$ according to Eq.
~\ref{eq:melting_curve}.
 }
    \label{fig:melt_composition}
\end{figure*}

Next, we have analysed the local geometric motif in the NPs using the centrosymmetry parameter\cite{centrosymm}.    
In addition we have classified the atomic environments using Polyhedral Template Matching\cite{P5714} (PHTM), which reveals stacking faults inside most of the NPs. Most atoms are located in an FCC environment separated by grain boundaries with local HCP order originating from atoms in intersecting regions of two surface facets with different orientations leading to a relative shift of two FCC grain regions. More precisely, we find that the grain boundaries meet at the NP vertices dividing the FCC environments into pyramidal regions. Similar observations have also been made for Au-Pd NP simulations~\cite{Mejia-Rosales2007}. Only in a few cases, such grain boundaries are completely absent,  e.g. for $x_{\text{Zn}}=20~\%$. 
\\
Moreover, we investigate the temperature $T_{\text{Lat}}$ at which an ordered lattice is formed in the SA cooling process by analyzing the NP structure at each $T$ using PHTM. In addition, the composition dependent melting points $T_m$ have been determined by extracting the turning point from Lindemann crystal parameters\cite{Lindemann1910} as a function of $T$\cite{JanPaper1}. For both quantities $T_{\text{Lat}}$ and $T_m$ (s. Fig.~\ref{fig:melt_composition}) we find a linear dependence on $x_{\text{Zn}}$.
Consequently, brass NPs with a higher Zn content must be cooled to a lower temperature to adopt an ordered lattice, which is also in agreement with the decreasing melting temperature of bulk brass with increasing Zn contents \cite{brass}. In addition, $T_{\text{Lat}}$ is found to be approximately $\SI{250}{\kelvin}$ lower than the melting temperatures of the NPs with the same composition (s. Fig.~\ref{fig:melt_composition}) indicating a hysteresis-effect, which might depend to some extent on the simulation protocol. Still, as discussed in Ref.~\citenum{C6CP08606C} the crystallization temperature and melting temperatures of NPs can differ several hundred Kelvin. The linear decrease of $T_m$ with the Zn atom fraction can be attributed to the increase, i.e. less negative value, of the cohesive energy with $x_{\text{Zn}}$, and consequently less energy is required for the melting process \cite{Attarian_Shandiz_2008}. 

A comparison between the composition-dependent melting coefficient $k_c$ of bulk $ \alpha$-brass $k_c =  (5.34 \pm 0.1) ~\si{\kelvin}$ extracted from the phase diagram\cite{brass} by linear regression and of the coefficient for the 1103 atom NPs, $k_c =  (5.37 \pm 0.2) ~\si{\kelvin} $ shows a very good agreement. Since the composition dependence of $T_m$ for brass NPs is apparently very similar to the bulk phase, the most significant difference of $T_m$ is the size dependent intercept value, scaling like $T_{m} (N)=  T^{\text{B}}_{m} \cdot (1 -   k_N \cdot N^{-\frac{1}{3}})$ where $T^{\text{B}}_{m}$ is the bulk Cu melting temperature\cite{JanPaper1}. This suggests that the melting temperatures of $\alpha$-brass NPs can be approximated by

\begin{align}
T_{m}^{\text{mod}} =  T^{\text{B}}_{m} \cdot  (1 -   k_N \cdot N^{-\frac{1}{3}}) - 100 \cdot k_c  x_{\text{Zn}}~,
\label{eq:melting_curve}
\end{align}

where $k_N$ and $k_c$ are size and composition dependent melting coefficients.

Preliminary tests using a smaller $N=165$ atom nanoparticle indicate that the composition dependent melting constant $k_c$ seems to be independent of the size of the nanoparticle. The results of these additional melting point investigations are outlined in the SI, Sec.~1.2.3.

Values for $k_N$ and $T^{\text{B}}_{m}$ extracted from simulations with the HDNNP can be calculated from the size-dependent melting curves of Wulff-shaped\cite{Wulff1901} or spherical Cu NPs as shown in our previous work \cite{JanPaper1}. We find that $k_N$ is between $2.44~\pm~0.1$ and $2.5~\pm~0.1$ and $T^{\text{B}}_{m}$ is between $\SI[separate-uncertainty = true]{1355(13)}{\kelvin}$ and $\SI[separate-uncertainty = true]{1299(15)}{\kelvin}$ for Wulff-shaped and spherical NPs, respectively. The melting curve predicted by Eq.~(\ref{eq:melting_curve}) is also shown in Fig.~\ref{fig:melt_composition}a, the error bars originate from the statistical uncertainties of $T_{m}^{\text{Wulff},~\text{B}}$ and $k_{N}^{\text{Wulff}}$, obtained by linear regression as discussed in Ref.
~\citenum{JanPaper1}. The size and composition dependent melting phase diagram as described by Eq.~(\ref{eq:melting_curve}) for large Wulff-shape $\alpha$-brass NPs with up to $N=100000$ atoms corresponding to an approximate diameter of $\SI{12}{\nano \meter}$ is shown in Fig.~\ref{fig:melt_composition}b.

\section{Conclusion} \label{sec:conclusion}

In this work we have investigated the relationship between size, composition, element-specific site occupations and the shape of large $\alpha$-brass NPs containing thousands of atoms. Since these systems are too large for a direct application of electronic structure calculations like DFT, a high-dimensional neural network potential has been employed to provide the energies and forces with first-principles accuracy. 
\\
Monte Carlo simulations in the Semi-Grand Canonical Ensemble and simulated annealing molecular dynamics simulations have shown that with increasing zinc chemical potential zinc atoms first accumulate in the outermost atomic layers, which is particularly pronounced in small nanoparticles due to the high ratio of surface to bulk atoms. Only very high zinc contents lead to alloy formation in the interior of the particles, and in some cases the emerging structures could be identified as known bulk phases of $\alpha$-brass. Further, we found that the Zn concentration and distribution depends strongly on temperature. Grain boundaries caused by differently oriented surfaces divide the NPs to regions with local FCC order where the shape of brass NPs depends on the composition, which is particularly interesting for applications in catalysis, as surface defects resulting from internal grain boundaries have been associated with catalytically active sites \cite{P3221,P5906}. 

\section{Supporting Information}
Contains details about elemental distributions in brass surfaces, temperature dependence of the composition and elemental distributions and about the implications of combined SCGE and relaxation simulations.

\begin{acknowledgement}
We thank the Deutsche Forschungsgemeinschaft (DFG) for financial support (Be3264/10-1, project number 289217282 and INST186/1294-1 FUGG, project number 405832858). We would also like to thank the North-German Supercomputing Alliance (HLRN) under project number nic00046 for computing time. 
\end{acknowledgement}

\bibliography{main}


\end{document}